\documentstyle[12pt]{article}
\newcommand{\be}{\begin{equation}}
\newcommand{\ee}{\end{equation}}
\newcommand{\ba}{\begin{eqnarray}}
\newcommand{\ea}{\end{eqnarray}}

\title{
Fractional dimensional Fock space and Haldane's exclusion statistics.
q/p case. }
\author{
K.N. Ilinski $^{1,2}$\thanks{E-mail: kni@th.ph.bham.ac.uk} $\quad$
J.M.F. Gunn$ ^{1}$,\thanks{E-mail: jmfg@th.ph.bham.ac.uk} \\
{\small\it $^{1}$ School of Physics and Space Research,
University of Birmingham,}
\\
{\small\it Birmingham B15 2TT, United Kingdom} \\
[0.5cm]
{\small\it $^{2}$ Institute of
Spectroscopy, Russian Academy of Sciences,} \\
{\small\it Troitsk, Moscow region,
142092, Russian Federation}}

\date{  }

\begin{document}
\setcounter{page}{0}
\maketitle
\vskip -9.5cm
\vskip 9.5cm
\thispagestyle{empty}
\begin{abstract}
The discussion of Fractional dimensional Hilbert spaces in the
context of Haldane exclusion statistics is extended from the case \cite{IG} of
$g=1/p$ for the statistical parameter to the case of rational $g=q/p$ with
$q,p$-coprime positive integers.
The corresponding statistical mechanics for a gas of such particles is
constructed. This procedure is used to define the statistical mechanics for
particles with irrational $g$. Applications to strongly correlated systems such
as the Hubbard and $t-J$ models are discussed.
\end{abstract}
cond-mat/9506115\break
\vspace{1cm}

\newpage

\section{Introduction}
Last year a series of papers devoted to Haldane exclusion statistics and
a generalized Pauli principle have appeared. Two new concepts
were introduced in Haldane's original paper~\cite{Hald}.
Let us briefly review them.
\begin{enumerate}
\item The first concept
is the {\it Haldane dimension}, $d(N)$, which is the dimension of
the one-particle Hilbert space associated with the $N$-th particle,
keeping the coordinates of the other $N-1$ particles fixed.
The statistical parameter, $g$, of a particle (or `$g$-on') is then defined
by the equation
\be
g = - \frac{d(N+m) - d(N)}{m}
\label{g}
\ee
and conditions of
homogeneity on $N$ and $m$ are imposed. In addition the system is confined
to a finite region where the number $K$ of independent single-particle
states is finite and fixed.  Here the usual Bose and Fermi ideal gases have
$g=0$ for Bose case (i.e. $d(N)$ does not depend on $N$)
and $g=1$ for Fermi case -- that is the dimension is reduced by unity
for each added fermion, which is the usual Pauli principle.
\item
The second concept is the {\it Haldane-Wu state-counting procedure}, which is
a combinatorial expression for the number of ways,
$W$, to place $N$ $g$-ons into $K$ single--particle states.
\be
W = \frac{(d(N)+N-1)!}{(d(N)-1)! N!}  \qquad d(N)=K-g(N-1) ,
\label{W}
\ee
This expression was used by many authors \cite{Wu,WB,W2,Raj} to
describe the thermodynamical properties of $g$-ons. In particular, in Refs.
\cite{WB,MS2,Ha,Is} it was shown that excitations in
the Calogero--Sutherland model obey the form of statistical mechanics defined
by Wu
\cite{Wu} for $g$-ons, with fractional $g$, in general.
(Possible generalizations, which depend linearly on $g$, of the Haldane-Wu
state-counting procedure were considered in Ref.~\cite{IIG}).
\end{enumerate}

In our previous paper~\cite{IG}
we argued that these two ideas introduced
by Haldane are {\it distinct}, in that the most natural combinatorial
expression for $W$, using the quantity $d(N)$, is {\it not} (\ref{W}).
This led us to a definition of the dimension of a Hilbert space which
is {\it fractional} and is related in a natural manner to particles having
a fractional $g=1/m$. We constructed the statistical mechanics for such
particles and showed that many features agree with Wu's statistical
mechanics, however the agreement is not complete. Since these arguments are
relevant to this letter, let us briefly summarise them.

The first argument for the expression for $W$ in (\ref{W}) is motivated by
initially remembering the fermionic case, writing (\ref{W}) in the form
$$
W_{F} = K (K-1) (K-2) (K-3) \cdots (K-N+1) / N!
$$
where the $i$th bracket in the numerator is the number of ways to
insert the $i$th particle into the system when all the particles are
distinguishable. The answer is then corrected, for indistinguishability, by the
$N!$ in the
denominator. By analogy, expression (\ref{W}) can be rewritten replacing
the factor of the {\it initial} number of available states, $K$, in
fermionic expression by an effective number of allowed states,
$K+(1-g)(N-1)$. Thus the space of available single-particle states {\it
swells} before the particles are added.  This contradicts the assumption of
fixing the size of the system which leads to a fixed number of
single-particle states. From the definition of the Haldane dimension,
$d(N)$, we would have na\"{\i}vely expected a result of the form
$$
W = K
(K -g)\cdots(K+(1-g)(N-1)-N+1) / N!
$$
where the number of available
states decreases in a manner proportional to $g$ as the particles are added.
This
expression has the disadvantage that it does not even give us the
correct interpolation to the Bose limit.

Another argument (which is perhaps more relevant to the subject of this
paper) arises from comparing the prediction of
(\ref{W}) for the case $g=2$ and a straightforward calculation for $K$
single-particle states in the $N$-particle sector. For example with $K=10$ and
$N=3$ the Haldane-Wu procedure gives $W=8!/(3!5!) =56$ and straightforward
calculation gives $W_{0}=(10 \cdot 8 \cdot 6)/3!=80$. It is easy to see that,
for general large $K$ and $N$, the deviation of $\ln W$ from straightforward
counting is the following:
\be
\ln W_{0} - \ln W = \frac{1}{2}\frac{N^2}{K} +
O\left(\frac{N^3}{K^2}\right) \ .
\ee
This deviation is important for
cases where the occupation numbers are not small.  So, the above discussion
may be summarised by stating that Haldane's definition of the fractional
dimension and the Haldane-Wu state-counting procedure are not consistent.
Moreover, {\it we cannot expect the agreement between predictions of
theories based on $W$ and state-counting procedures, which fit $W_{0}$, for
$g=2$}.

Certainly one may ask why one should consider state-counting
procedures such as $W_{0}$. Many applications of $W$ are known, so it might be
said that one should regard the Haldane dimension $d(N)$ as a less important
variable. However, our main motivation in considering the original
Haldane definition of exclusion statistics is to note their realization in
strongly correlated systems such as the Hubbard model
with infinite $U$ (with statistical parameter $g=2$) or the $t-J$ model.
There the charge degree of freedom on one site exactly obeys the $W_{0}$
state-counting procedure with $g=2$.
Examples for the other integeral statistical parameters are readily constructed
using fermions with more spin states or `flavours'. Another possibility is to
model finite (as against infinite) interaction effects by varying the
statistical parameter, $g$, from $2$ to $1$.

These problems require an
appropriate definition of: $d(N)$, a fractional dimensional Hilbert space and
the state-counting procedure corresponding to $W_{0}$ for general rational
$g=q/p$ (In the paper~\cite{IG} only the simplest case $g=1/m$ was considered).
These issues will be addressed in this paper. We will see that the agreement
with
statistical mechanics based on $W$ (\ref{W}) is decreases
with increasing $q$ but this is not surprising because of the aforementioned
disagreement of the state-counting procedures for integer $g$.

In the next sections we generalise the considerations of
Ref.~\cite{IG} for $g=q/p$. We will give the description of the corresponding
Hilbert space, a calculation of dimension which reproduces the correct Haldane
dimension $d(N)$ and discuss the corresponding statistical mechanics.

\section{Hilbert space definition of Haldane's dimension}

In this section we review the Hilbert space definition of Haldane
dimension for fractional values of statistical parameter $g$. To do this we
follow Ref.~\cite{IG}.

First of all let us return to the motivation for
the introduction of Haldane's dimension, $d(N)$, and discuss an alternative
view of this quantity. In the original definition the dimension $d(N)$
reflects the number of independent possibilities to add the $N$-th particle to
the system.  Intuitively it is clearest to consider a system on a lattice.
The decrease
in the number of allowed states $d(N)$ with the increase in the number of
particles is a consequence of the exclusion of the particles and gives us
{\it a Generalized Pauli exclusion principle}.  Haldane's suggestion, to
complete such a statistical description, was to impose linearity of the
variation in $d$ as a function of $N$. The coefficient of proportionality is
$g$.  This
immediately implies an obvious condition on the statistical constant $g$:
the allowed number of states for the first particle is equal to the number of
independent degrees of freedom of the system,
$K^{\prime}$, and to keep the particle homogeneity condition we have to
assume that $g=K^{\prime}/M$ where $M$ is the maximum allowed number of
particles (i.e. $g$ should be rational).  Moreover, if we assume that the
statistical parameter does not
depend on the number $K^{\prime}$ (and statistics can be realized on single
site with $K^{\prime}=1$), we must only consider the set of statistical
parameters
$$
  g = \frac{1}{m} \qquad m = 1,2,3,... \in N \ .
$$
This case was considered in Ref.\cite{IG}.
To consider general fractional values (or indeed $g$ integral and greater than
unity) of statistical parameter, $g=q/p$ (where $q$ and $p$ are coprime
positive integers), we must relax the condition of $g$ being independent of
$K^{\prime}$, and allow use of $K^{\prime}>1$. To describe $g=p/q$, we require
at least $q$ of them (for example for the $t-J$ model
we have two single-particle levels per each site). Generally,
the set of states should be divided into blocks with $q$ states
in each of them and the number of
states $K^{\prime}=qK$.

 Our main idea now is to consider
the process of inserting of the $N$-th particle into the system as a
probabilistic process (in Gibbs spirit), i.e. we assume that the
probability of such an insertion plays the role of the Haldane's measure of
the probability to add the $N$-th particle to the system. Let us illustrate
the idea for the case of a single block (with $q$ levels) and give an
interpretation of $d(N)$ in that case.

First of all we have the vacuum state (empty block states) to which we can
add the first particle. We assume that the statistics reveals itself at the
level of two particles and it is irrelevant for the case $N=1$,  so
$d(1)=q$ (because we can put the particle in any of $q$ levels in the block).
Now let us assume that we cannot add the second particle
to the system at {\it each attempt} and the process is a probabilistic one
with the probability $(1-1/p)$ of success per each level. Then result
probability to add the second particle to the block is
$q(1-1/p)$. Then if we consider a large
number, $Q$, of copies of the system and perform trials with each,
we will find approximately $q(1-1/p)Q$ double-occupied blocks so that on
average only a $q(1-1/p)$ part of the particle is in each block. We interpret
this as a fractional dimension of the subspace with double occupation of
the block and $d(2)=q(1-1/p)$. This implies that the statistical parameter
will be $g=d(1)-d(2)=q/p$.

Let us continue the procedure and consider a third particle. We can repeat the
previous argument with only a small correction: now we
consider the conditional probability to add a third particle to the systems
with the condition that there are already two particles in the system (whose
coordinates we have to fix) before the trials.  This conditional
probability is, by assumption, $(1-2/p)$ per level which determines the
average probability for the third particle in each block and the equality
$d(3)=q(1-2/p)$.  Then the full probability to find the block with $N=3$ is
$q^{2}(1-1/p)(1-2/p)$.

We see that the probability to find $N>p$ particles in the block is equal to
zero and this avoids difficulties with a
nonpositive number of $N$-particle states which occurs in the approach based on
the expression (\ref{W})~\cite{W2,Pol}. This is achieved by
eliminating of the probability on the $m+1$-th step. In this context the
original Pauli principle `There are no double-occupied states'
can be reformulated as `The probability to find a double-occupied state is
equal zero'.

Moreover we can
give a `geometrical' definition of the fractional dimension by drawing
parallels with the notion of a noninteger dimension in the framework of
dimensional regularization. Indeed, usually tin the calculation of
thermodynamical
quantities of an ideal gas, such as the partition function or the mean value of
an
arbitrary physical variable $\hat{O}$, we have to compute the following
traces:
\be
Z=Tr (Id \cdot e^{-\beta H}) \qquad \mbox{or} \qquad
\langle\hat{O}\rangle=Tr (Id \cdot e^{-\beta H} \cdot \hat{O})
\label{Z}
\ee
where the Hamiltonian $H$:
\be
H = \sum_{i,j=1...K,q} \epsilon _{i} n_{i,j}
\label{Ham}
\ee
is of the usual ideal gas form, and does not depend on the statistics but
depends
only on the occupation numbers of the particles on the $i$-th block and
an `unit operator' $Id$; these completely define the exclusion statistics of
the particles:
\be
Id = \sum_{\{n_{i,j}\geq 0\}} \alpha_{\{n_{i,j}\}_{i,j=1}^{K,q}}
|\{n_{i,j}\}_{i,j=1}^{K,q}\rangle\langle \{n_{i,j}\}_{i,j=1}^{K,q}| \ .
\label{Id}
\ee
Here $|\{n_{i,j}\}_{i,j=1}^{K,q}\rangle$ is the state with $n_{i,j}$ particles
in
the state $i,j$, and
$\alpha_{\{n_{i,j}\}_{i,j=1}^{K,q}}$ is the probability to find this state.
It is obvious that the expression (\ref{Z}) gives correct answers for
fermions and bosons and there is no contradiction in using the expression for
intermediate statistics. Moreover we can say that the
Hilbert space of the theory is constructed if we define the operator $Id$
and use the equalities (\ref{Z}) (the scalar product can be defined in the same
way). We will not go more into the mathematical
details here and will discuss them elsewhere.

We can now interpret  $\alpha_{\{n_{i,j}\}_{i,j=1}^{K,q}}$
as the dimension of
the subspace spanned by the vector $|\{n_{i,j}\}_{i,j=1}^{K,q}\rangle$.
Indeed, usually the
dimension of a subspace $S$ can be defined as the trace of unit operator on
the subspace
$$
\dim S = Tr(Id|_{S})
$$
(this definition was used in dimensional regularization where $d-\epsilon =
\sum_{i} \delta _{i}^{i}$). The full dimension of the $N$- particle
subspace of the space of states is then given by the formula:
\be
W_{0}= \sum_{\{n_{i,j}\geq 0;\sum n_{i,j}=N\}}
\alpha_{\{n_{i,j}\}_{i,j=1}^{K,q}} = Tr (Id|_{\sum n_{i,j}=N})
\label{W01}
\ee
which we will use later for the state-counting procedure in the next
section. Haldane's dimension $d(N)$ of the $N$-th particle subspace
with an arbitrary fixed $N-1$-particle substate
$|\{n_{i,j}\}_{i,j=1}^{K,q}\rangle, \sum n_{i,j}=N-1$ is then described by the
relation:
\be
d(N)= \sum_{i,j}
\frac{\alpha_{\{n_{i,j}\}_{i,j=1}^{K,q}}}
{\alpha_{\{n_{i,j}\}_{i,j=1}^{K,q}}|_{\sum n_{i,j}=N-1}} \ .
\label{dalph}
\ee
(This is a sum of conventional probabilities to add $N$-th particle to the
system with the condition that before the addition the system is in the state
$|\{n_{i,j}\}_{i,j=1}^{K,q}\rangle,\sum n_{i,j}=N-1$).

The statements made so far are general and did not require any concrete choice
of the
probabilities $\alpha_{\{n_{i,j}\}_{i,j=1}^{K,q}}$. Moreover when we
discussed these ideas for the single block we defined
probabilities with only a single index and so we have a choice to
define probabilities with several indices.

In fact, there is
single self-consistent way to define $\alpha_{\{n_{i,j}\}_{i,j=1}^{K,q}}$ such
that
\begin{enumerate}
\item  the definition of the $N$-th particle dimension $d(N)$
(\ref{dalph}) actually gives Haldane's conjecture $K-g(N-1)$ for
$d(N)$;
\item The Hilbert space of the system with $K$ blocks is
factorized into the product of Hilbert spaces corresponding to each block.
This property together with the Hamiltonian of the form
(\ref{Ham}) is characteristic of an ideal gas of the particles with any
statistics and will lead to the factorization of partition function
and other physical quantities.
\end{enumerate}

To prove this let us use the assumption about the tensor
product nature of the full Hilbert space which immediately implies the
operator $Id$ for the complete system is a tensor product of the operators
$\{Id_{i}\}_{i=1}^{K}$ for each state:
\be
Id = Id_{1}\otimes Id_{2}
\otimes \cdots \otimes Id_{K} \ .
\label{TP}
\ee
The last relation is
equivalent to the following expression for probabilities
$\alpha_{\{n_{i,j}\}_{i,j=1}^{K,q}}$:
\be
\alpha_{\{n_{i,j}\}_{i,j=1}^{K,q}}= \prod_{i=1}^{K}
(1-1/p)(1-2/p)...(1-(\sum_{j} n_{i,j}-1)1/p) \ .
\label{alph}
\ee
Combined with
(\ref{dalph}) we obtain the required equality for $d(N)$:
\be
d(N)=
\sum_{l=1,j=1}^{K,q} (1-1/p \sum_{j^{\prime}}n_{l,j^{\prime}})
= q(K - 1/p(N-1)) \ .
\label{Hal}
\ee
Moreover, we can say that
$W_{0}$ in (\ref{W01}) together with Haldane's dimension
$d(N)$~(\ref{Hal}) and the corresponding $\alpha_{\{n_{i,j}\}_{i,j=1}^{K,q}}$
(\ref{alph}) replaces expression (\ref{W}) in the theory. These are used in the
investigation of thermodynamical properties in the rest of the paper.

We conclude this section with a note about the connection between the subject
of section 2 and~Ref\cite{Pol}, where the microscopic origin of the
Haldane-Wu state-counting procedure was examined. As in this work, the
notion of statistics was considered in a probabilistic spirit.  The author
assumed that a single level may be occupied by any number of particles, and
each occupancy is associated with an a priori probability. These
probabilities are determined by enforcing consistency with {\it the
Haldane-Wu state-counting procedure} and not with {Haldane's definition
of exclusion statistics}.  In detail, the distinction with the current work
are: the definition of the Haldane dimension was not used, and a Hilbert space
was not constructed and the a priori probabilities in Ref \cite{Pol} may be
negative. As we have shown\cite{IG}, the Haldane dimension and the
Haldane-Wu state-counting procedure are distinct; the formulations of this work
and Ref.\cite{Pol} are correspondingly distinct.

\section{Statistical mechanics of $q/p$-ons}

Let us consider initially the second virial coefficient, $B_{2}$, which
reflects the statistical interactions between the particles and has been
calculated for the other representations of particles obeying exclusion
statistics \cite{Wu,MS2}
$$
B_{2} = - Z_{1}
(2 Z_{2}/Z_{1}^{2} - 1) .
$$
where $Z_{i}$ is the $i$-particle's partition function.
For the case of $g=q/p$, a straightforward calculation gives:
$$
 Z_{2} = e^{-2\beta \epsilon} (q^2 C^{2}_{K} + K \frac{q(q+1)}{2}(1-1/p)) \ ,
\qquad
 Z_{1} = e^{-\beta \epsilon} qK
$$
which results in
\be
B_{2} = q/p + 1/p - 1
\label{Vir}
\ee
coinciding with Wu's expression \cite{Wu,MS2}, $B_{2} = 2q/p - 1$, only for the
case $q=1$, but yields the expressions which might be anticipated, on physical
grounds, for integer values of $g$ where Wu's result is more difficult to
interpret.

To calculate the partition function of an ideal $g$-on gas, we return to the
formulae (\ref{Z}),(\ref{Ham}) of the previous section and use our
self-consistent choice of the operator $Id$ (\ref{Id}) with the
coefficients (\ref{alph}). As was noted above, such a choice allows us to
factorize the statistical operator $Id$ into a tensor product of the
operators for single states (\ref{TP}) which leads to the factorization of
the partition function of the system:
\be
Z = \prod _{i=1}^{K} Z_{i}  \ .
\label{PF}
\ee
Here the function $Z_{i}$ is the partition function associated with a
single block labelled by index $i$:
$$
Z_{i} = Tr(Id_{i} \cdot e^{-\beta (\epsilon_{i} - \mu)n_{i}}) \ .
$$
For the case of $g=q/p$ the last expression can be rewritten as
$$
Z_{i}=1+qe^{-\beta (\epsilon_{i}-\mu)}
+\frac{q(p-1)(q+1)}{2p}e^{-2\beta (\epsilon_{i}-\mu)}
\cdots+\frac{p(p-1)(p-2)...1}{p^{p}}\ C^{p}_{q+p-1}\
e^{-p\beta (\epsilon_{i}-\mu)}
$$
or in a more closed form
\be
Z_{i}=\sum_{n=0}^{p} \ \ C^{n}_{p} \ \ C^{n}_{q+n-1} \ \ n! \ \ \frac{e^{-n
\beta
(\epsilon_{i}-\mu)}}{p^{n}} \ .
\label{Zi}
\ee
It is obvious that the formulae (\ref{PF}), (\ref{Zi}) interpolate
between integer $g$ (in particular, Fermi ($g=p=1$)) and Bose ($g=0,p=\infty$)
cases and hence lead to
the an interpolation for all statistical quantities.

The most interesting object for comparison (for different $p$ and $q$) is the
distribution function which in our approach may be calculated as (we once
more put all energies equal to the single $\epsilon$ to make the
expression more compact)
$$
n(\beta ,\mu) \equiv \langle N/qK\rangle = \frac{1}{\beta}
\frac{\partial \ln Z_{i}}{\partial \mu}
$$
which results in the next equality:
\be
n(\beta ,\mu) = 1/q
\left(\sum_{n=1}^{m} \ C^{n}_{p}\ C^{n}_{q+n-1} \ n! \ n \
\frac{e^{-n \beta (\epsilon_{i}-\mu)}}{p^{n}}\right) \cdot
\left(\sum_{n=0}^{m} \ C^{n}_{p} \ C^{n}_{q+n-1} \ n! \ \frac{e^{-n \beta
(\epsilon_{i}-\mu)}}{p^{n}}\right)^{-1} \ .
\label{n}
\ee
Let us briefly discuss the behaviour of this distribution function
at low temperature (in the high temperature limit all statistical
effects disappear).
We can easily see from (\ref{n}) that in the low
temperature limit the function $n(\beta ,\mu)$ for our $q/p$-particles
behaves exactly like the standard distribution function \cite{Wu,Ouvry}, i.e.
at any positive value of $g$ we have a `Fermi level':
\be
\begin{array}{ll}
n(\beta ,\mu)= p/q  & \qquad \mbox{for} \qquad  \epsilon < \mu   \\
n(\beta ,\mu)= 0 \ \   & \qquad \mbox{for} \qquad  \epsilon > \mu
\end{array}
\nonumber
\ee
At low,
but finite, temperatures we
will have a discrepancy with the standard result. To simplify the comparison we
 remember that the
distribution function of $g$-ons in the Haldane-Wu approach can be expressed
as
\be
n(\beta ,\mu) = \frac{1}{w(\beta ,\mu) + g}
\label{w}
\ee
and the function $w(\beta ,\mu)\equiv w(\xi)$ is a solution of
equation~\cite{Wu,Ouvry}:
\be
w^{g}(\xi) (w(\xi)+1)^{1-g} = e^{\beta (\epsilon-\mu)}\equiv \xi \ .
\label{w1}
\ee
At low enough, but finite, temperatures and energy $\epsilon$ above
the Fermi level (i.e. $\xi $ is very big) we find the following
expansion for $n$ as a function of variable $\xi$:
\be
n(\beta ,\mu) = \frac{1}{\xi}(1 + \frac{1}{\xi} (1-2g) + ... ) \ .
\label{a}
\ee
The first term on the RHS is a pure Boltzmann distribution and this is
a common feature for any statistics in this limit. But the next term
reflects the statistics of the particles. There is no difficulty in showing
that in the same limit expression~(\ref{n}) gives us similar
asymptotics for the distribution function:
\be
n(\beta ,\mu) =
\frac{1}{\xi} (1 + \frac{1}{\xi} (1-q/p-1/p) + \cdots ) \ ,
\label{Ex1}
\ee
that is consistent with comparison of second virial coefficients and coincides
with (\ref{a}) for the case $q=1$.

Below the Fermi level at low, but finite,
temperatures $\xi$ is a small, but finite, parameter. From
eq.(\ref{w1}) we immediately obtain the following approximation for
$n(\xi)$:
$$
n(\beta ,\mu)  = \frac{p}{q} - \frac{p^{2}}{q^{2}}\xi ^{p/q} + \cdots  \  \ .
$$
In contrast with this the asymptotic behaviour for the distribution function
(\ref{n})
contains terms which are linear in $\xi$:
\be
n(\beta ,\mu) = \frac{p}{q}\left(1 - \xi \frac{p}{q+p-1} +\cdots  \  \ \right)
{}.
\label{Ex2}
\ee

In conclusion of this section let us return to the state-counting procedure
for probabilistic $g$-ons.
For convenience set all energies to be identical $\epsilon_{i}=\epsilon$ and
rewrite the formulae for the dimension of the $N$ -particle subspace of the
full Hilbert space (\ref{W01}) in the form
without constrained summation:
$$
W_{0}= \sum_{n_{i,j}=0}^{\infty}
\alpha_{\{n(i,j)\}_{i,j=1}^{K,q}} \cdot \delta_{{\scriptstyle
\sum}_{i,j}n_{i,j},N}
$$
which then will be treated as a Fourier integral:
$$
W_{0}= \frac{1}{2\pi}
\int _{0}^{2\pi } d\phi \sum_{n_{i,j}=0}^{\infty}
\alpha_{\{n(i,j)\}_{i,j=1}^{K,q}}\cdot e^{i \sum_{i,j=1}^{K,q} n(i,j)
-N)\phi} \ .
$$
This last
equality and the factorization property (\ref{alph}) for the coefficients
$\alpha$ allow us to perform the summation over
$n_{i,j}$ and obtain an expression for $W_{0}$ with only a single
integral over the auxiliary variable $\phi$:
\be
W_{0}= \frac{1}{2\pi} \int _{0}^{2\pi } d\phi \ \
z(e^{i \phi})^K \cdot e^{-i N\phi} \ .
\label{int}
\ee
where the function $z(\exp ^{i \phi})$ coincides with the single state
partition function for imaginary energy and chemical potential ($\beta (\mu-
\epsilon) \rightarrow i \phi $):
$$
z(e^{i \phi})=\sum_{n=0}^{p} \ \
C^{n}_{p} \ \ C^{n}_{q+n-1} \ \ n! \ \ \frac{e^{i n \phi }}{p^{n}} \ .
$$
Eq.(\ref{int})
replaces the Haldane-Wu state-counting expression~(\ref{W}) in the theory
of probabilistic $g$-ons.

It is interesting to note that in some sense our consideration is complementary
to the Haldane-Wu case: we can easily find the distribution
function analytically but to find the entropy in the large $N$ limit an
algebraic equation of the $p$-th order must be solved, whilst in the Haldane-Wu
approach the
form of the entropy is obvious but an algebraic equation of the $p$-th order
is encountered when deriving the distribution function.

\section{Discussion}

Let us discuss a few of the results and their consequences in more detail.
Perhaps the least attractive aspect of our formulation is that the results
depend on both of the parameters $p$ and $q$. However this apparent drawback is
not as significant as one might initially presume: this is because as $p$ and
$q$ both become large, with their ratio fixed, the results only depend on
$p/q$. This may be seen by examining expressions
(\ref{Vir}),(\ref{Ex1}),(\ref{Ex2}).

An interesting extension of this logic is that we may define the statistical
mechanics of irrational $g$ by considering a sequence of rational numbers,
$p/q$, which approach the required irrational number.

In conclusion, the discussion of Fractional dimensional Hilbert spaces in the
context of Haldane exclusion statistics has been extended from the case
\cite{IG} of
$g=1/p$ for the statistical parameter to the case of rational $g=q/p$ with
$q,p$-coprime positive integers.
The corresponding statistical mechanics for a gas of such particles is
constructed.

\section*{Acknowledgments.}

We wish to thank A.S.Stepanenko for useful discussions.
This work was partially supported (K.N.I) by the Grant of the International
Science Foundation N R4T000, Grants of Russian Fund of Fundamental
Investigations N 94-02-03712 and N 95-01-00548, Euler stipend of German
Mathematical Society, INTAS-939 and by the UK EPSRC Grant GR/J35221.

\end{document}